# A Effective Carrier Phase Recovery Method in Tight Time-Packing Fast than Nyquist Optical Communication System

**Peng Sun, Xiaoguang Zhang\*, Dongwei Pan, Lixia Xi, Wenbo Zhang, Xianfeng Tang**
*State Key Laboratory of Information Photonics and Optical Communications, Beijing University of Posts and Telecommunications*
*Beijing, China*
*Author e-mail address: xgzhang@bupt.edu.cn\*, Pengsun@bupt.edu.cn*

**Abstract:** We propose a new scheme that combines polybinary transformation and corrected-BPS to compensate phase noise for PDM-FTN-QPSK when its accelerated factor is 0.5, which has 3.3 dB OSNR gain when phase noise is 800 kHz. © 2020 The Author(s)

## 1. Introduction

Faster-Than-Nyquist (FTN) is one of effective method to improve spectrum efficiency (SE), but its system inevitably brings inter-symbol interference ($ISI_{FTN}$), coming from the non-orthogonal design, that should be equalized [1]. The phase noise (PN) induced by the linewidth of the transmitter and local lasers should be tracked and be compensated whose procedure is called carrier phase recovery (CPR). Because of the issue of $ISI_{FTN}$ in the FTN system, which is not appeared in ordinary coherent system, the impairments equalization algorithms might be modified in order to be suitable to FTN system. For example, the conventional carrier phase recovery algorithms, such as Viterbi-Viterbi (V-V) algorithm and blind phase searching (BPS) algorithm generally cannot worked properly and should be modified.

In [2, 3], the authors have utilized modified $2^{nd}$-order digital phase-locked loop and optical domain pilot aided to recover carrier phase in FTN system, respectively. In recent years, polybinary transformation has attracted more attentions in FTN system [2, 4]. Its essential principle is to eliminate in-band noise by adding previous symbol to the present symbol [5]. In [6], the authors proposed a scheme that combine the polybinary transformation and pilot aided blind phase search (BPS) to eliminate the phase noise in FTN system. The aforementioned three schemes need certain overhead. Moreover, they are not applicable to the situation with a small time-squeezing factor where there is more serious $ISI_{FTN}$ in FTN system.

In this paper, we propose a new scheme that combines polybinary transformation and corrected-BPS which does not have pilot aided for PDM-FTN-QPSK system when its accelerated factor is lower and $ISI_{FTN}$ is more serious to compensate the PN.

## 2. Principle

In the FTN system, we can adjust time-squeezing factor $\alpha$ to compress bandwidth and improve SE, with the sacrifice of impairment of $ISI_{FTN}$. As depicted in Fig1. (a), when $\alpha$ = 0.5, the time packed QPSK shows 4 severely overlapped clusters in the constellation plane. Therefore, when we utilize conventional BPS algorithm to compensate PN, the conventional BPS algorithm will not effectively work for FTN system with low $\alpha$. Because during the stage of decision for conventional BPS algorithm, decision condition is reference from ideal constellation point of the Nyquist-QPSK whose amplitude of decision is {-1,1} and the number of its combination is 4. Therefore, the conventional BPS algorithm is not suitable to compensate the PN for FTN system when its $\alpha$ is lower and $ISI_{FTN}$ is more serious.

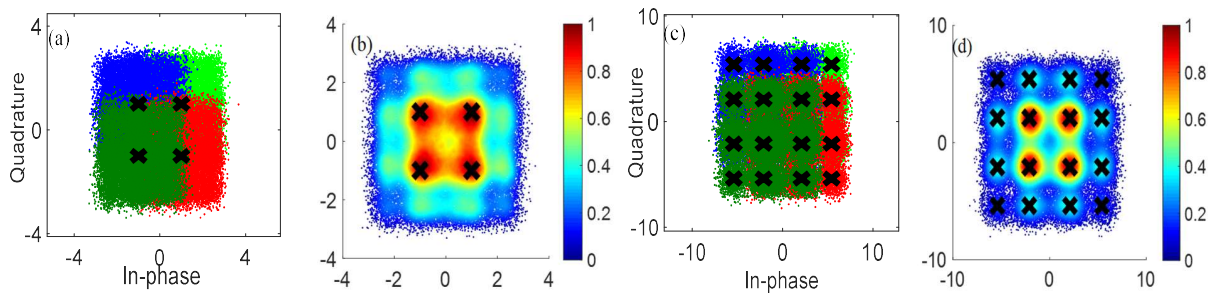

Fig.1. When the accelerated factor is equal to 0.5. (a) Constellation figure of the received signal. (b) Density distribution figure of the received signal. (c) Constellation figure of the received signal after polybinary transformation. (d) Density distribution figure of the received signal after polybinary transformation.

Fig1. (b) is the density distribution figure of Fig1. (a). We can ambiguously find 16 points, but their boundaries are difficult to distinguish. Faced with the above situation, we utilize polybinary transformation for received signal. Compared with Fig.1 (a), we still cannot find the effective information from Fig.1 (c) except amplitude is larger. However, as depicted in Fig.1 (d), the points are clearer than Fig.1 (b) and their amplitude of the points are {-5.4, -2.1, 2.1, 5.4}. According to the above phenomenon, we correct decision condition of BPS algorithm that the amplitude of decision change from {-1,1} into {-5.4, -2.1, 2.1, 5.4} and the number of point change from 4 into 16. By utilizing polybinary transformation and correcting decision condition for BPS algorithm, we can reuse BPS algorithm for FTN system to compensate the PN with lower $\alpha$. In contrast to the aforementioned three schemes [2,3,5], our new scheme does not need additional pilot aided and hence, still keep blind for BPS algorithm.

## 3. Simulation and Analysis

To verify the effectiveness of our new proposed scheme for FTN system with lower $\alpha$ and more serious $ISI_{FTN}$ to compensate the PN, we build a 28-GBaud PDM-FTN-QPSK simulation platform as shown in Fig. 2 (a). $\alpha$ and roll-off factor of root raised cosine filter are set to 0.5 and 0.1, respectively. In this system, we use MLSE algorithm to eliminate the $ISI_{FTN}$ and its number of tap is seven. We choose CMA algorithm for polarization de-multiplexing.

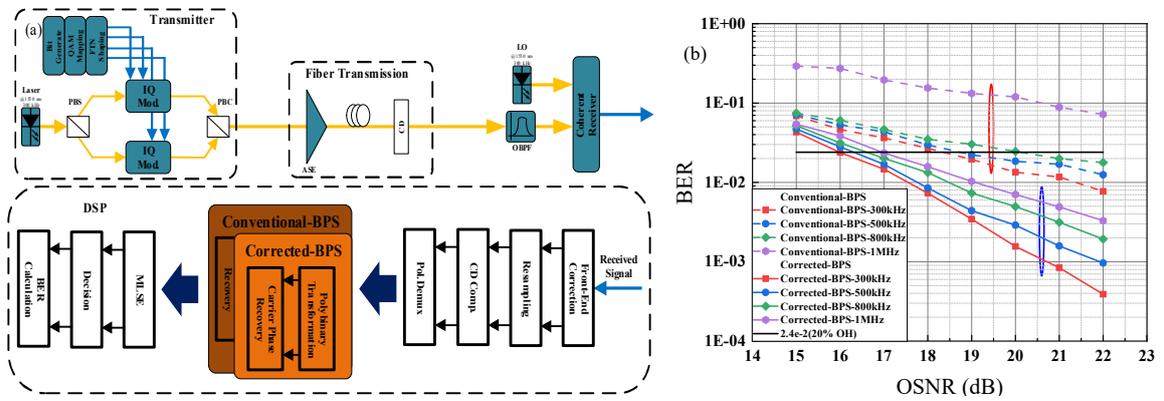

Fig.2. (a) Simulation platform. (b) BER vs. OSNR between conventional-BPS scheme and corrected-BPS scheme when PN are equal to 300 kHz, 500 kHz, 800klHz, 1 MHz, respectively.

The BER performance between the conventional-BPS scheme and our proposed corrected-BPS scheme when PN are respectively equal to 300 kHz, 500 kHz, 800 kHz and 1 MHz is depicted in Fig.2 (b). As shown in Fig.2 (b), we can clearly see that our proposed corrected-BPS scheme has better performance than the conventional-BPS scheme. When PN is equal to 300 kHz, 500 kHz and 800 kHz, our proposed scheme has achieved ONSR gain of about 2.3 dB, 2.6 dB and 3.3 dB compared to the conventional scheme, respectively.

## 4. Conclusion

In this paper, we propose a new scheme for PN compensation that combines polybinary transformation and corrected-BPS algorithm without pilot aided for PDM-FTN-QPSK system when its time-squeezing factor is lower and $ISI_{FTN}$ is more serious. The OSNR gain is about 2.3 dB, 2.6 dB and 3.3 dB compared with conventional scheme when PN are equal to 300 kHz, 500 kHz and 800 kHz, respectively.

This work was supported partly by the National Natural Science Foundation of China (61571057, 61527820, 61575082).